\documentclass[epjCONF, onecolumn]{svjour}
\usepackage{graphics}
\usepackage[varg]{txfonts} 
\usepackage[latin1]{inputenc}
\newcommand{\kms}{\rm\,km\,s^{-1}}
\newcommand{\kmskpc}{{\rm\,km\,s^{-1}{kpc}^{-1}}}
\newcommand{\pc}{\rm\,pc}
\newcommand{\kpc}{{\rm\,kpc}}

\newcommand{\degg}{{^\circ}}
\def\sun{{_\odot}}

\newcommand{\name}{{\tt PERLAS }}

\def\2s{2-$\sigma$}
\def\3s{3-$\sigma$}
\session-title{Conference Title, to be filled}
\begin{document}
\title{Kinematic groups across the MW disc: \\ insights from models and from the RAVE catalogue}
\author{T. Antoja\inst{1}\fnmsep\thanks{\email{antoja@astro.rug.nl}} \and A. Helmi\inst{1} \and F. Figueras \inst{2} \and M. Romero-G\'omez \inst{2} \and  RAVE collaboration}
\institute{Kapteyn Astronomical Institute, University of Groningen, PO Box 800, 9700 AV Groningen, the Netherlands\and Dept. d'Astronomia i Meteorologia, Institut de Ci\`encies del Cosmos (ICC), Universitat de Barcelona (IEEC-UB), Mart\'i Franqu\`es 1, E08028 Barcelona, Spain}
\abstract{
With the advent of the Gaia data, the unprecedented kinematic census of great part of the Milky Way disc will allow us to characterise the local kinematic groups and new groups in different disc neighbourhoods. First, we show here that the models predict a stellar kinematic response to the spiral arms and bar strongly dependent on disc position. For example, we find that the kinematic groups induced by the spiral arm models change significantly if one moves only $\sim0.6\kpc$ in galactocentric radius, but $\sim2 \kpc$ in azimuth. 
There are more and stronger groups as one approaches the spiral arms. Depending on the spiral pattern speed, the kinematic imprints are more intense in nearby vicinities or far from the Sun. 
Secondly, we present a preliminary study of the kinematic groups observed by RAVE. This sample will allow us, for the first time, to study the dependence on Galactic position of the (thin and thick) disc moving groups. In the solar neighbourhood, we find the same kinematics groups as detected in previous surveys, but now with better statistics and over a larger spatial volume around the Sun. This indicates that these structures are indeed large scale kinematic features.
} 
\maketitle
\section{Introduction}
\label{intro}

Sirius, Coma Berenices, Hyades, Pleiades, and Hercules are clear overdensities in the solar neighbourhood velocity distribution. We currently believe that all these groups may be due to the orbital effects of the non-axisymmetries of the Milky Way (MW) disc, that is, bar spiral \cite{Dehnen00,Quillen05,Antoja09}. However, groups such as the ones that are observed in the solar neighbourhood can be induced by different model parameter combinations. To break the degeneracy and use the kinematic groups to constrain the MW large scale structure, data from velocity distributions at different positions of the MW disc and predictions from the models are needed. 

In Section \ref{sec:1} we give some predictions on the behaviour of the kinematic structures generated by non-axisymmetric disc models in different disc positions. We focus on spiral arms effects as these have been less deeply studied than the effects of the bar. In Section \ref{sec:2} we show that with current data from RAVE we can leave the solar neighbourhood to study kinematic groups far from the sun, for the first time, and before the Gaia era. We end up with some conclusions (Section \ref{conc}).

\section{Models: spiral arm effects on the disc kinematics}
\label{sec:1}

\begin{table}
 \caption{Assumed values for the properties of the MW spiral arms in the present model.}
\label{tab:1}       
\centering
\begin{tabular}{llc}\hline
\hline\noalign{\smallskip}
 \multicolumn{2}{l}{Property}                        &Value or range   \\   \hline
\noalign{\smallskip}\hline\noalign{\smallskip}
Number of arms          &$m$                           & $2$             \\
Scale length            &$R_\Sigma$ ($\kpc$)           & $2.5$           \\
Locus beginning         &$R_{sp}$ ($\kpc$)             & $2.6$           \\
Pitch angle             &$i$  ($\degg$)                & $15.5$     \\
Relative spiral phase   &$\phi_{sp}(R_\sun)$ ($\degg$) & $88$    \\
Pattern speed           &$\Omega_{sp}$ ({$\kmskpc$})   & 20         \\  
Density contrast        &$K$                           & 1.6      \\  \hline
\noalign{\smallskip}\hline
 \end{tabular}
\end{table}

We perform test particles simulations with a gravitational potential for the MW which includes halo, bulge, disc, and spiral arms. The model for the spiral arms is called \name model \cite{Pichardo03}. This is a 3D mass distribution with more abrupt gravitational potential and forces than in the Tight-Winding Approximation (TWA, e.g. \cite{Bible08}). The model is tuned according to the observational ranges for the parameters of the spiral structure in the MW. 
The model, the initial conditions and an exploration of the whole free parameter space can be seen in \cite{Antoja11}. 

\begin{figure}
\centering 
\resizebox{0.35\columnwidth}{!}{%
 \includegraphics{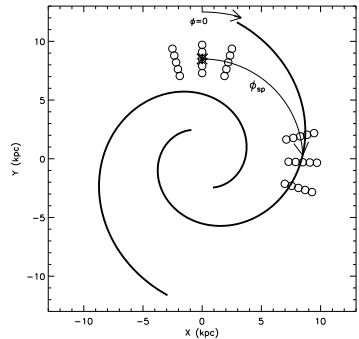} 
}
\caption{Adopted locus for our spiral arm model. The  assumed solar position is at $X = 0$ and $Y = 8.5 \kpc$. Open circles indicate several regions near the solar neighbourhood and near the spiral arm.}
\label{fig:1}       
\end{figure}

\begin{figure}
\centering
\resizebox{1.\columnwidth}{!}{%
  \includegraphics{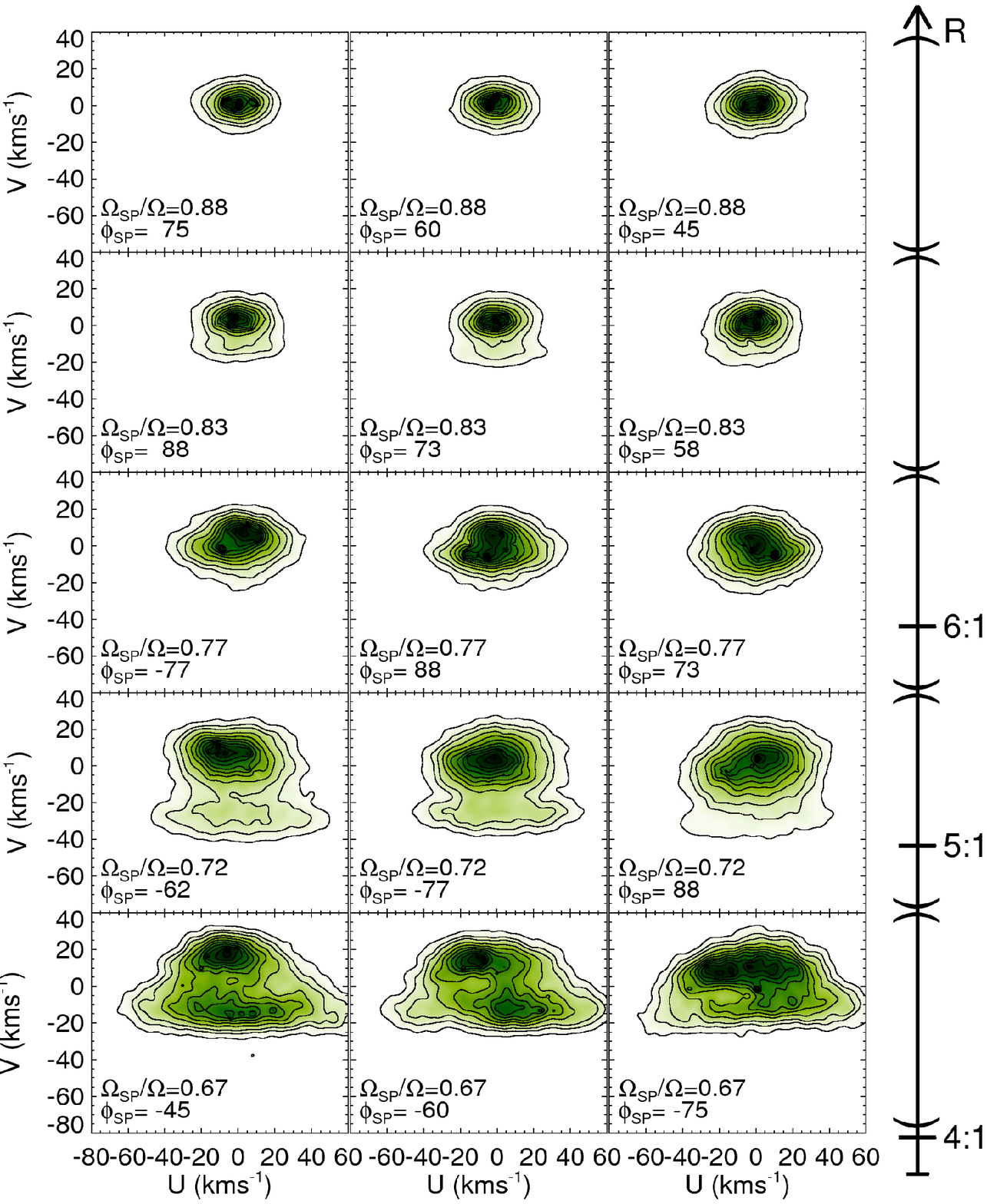} 
\includegraphics{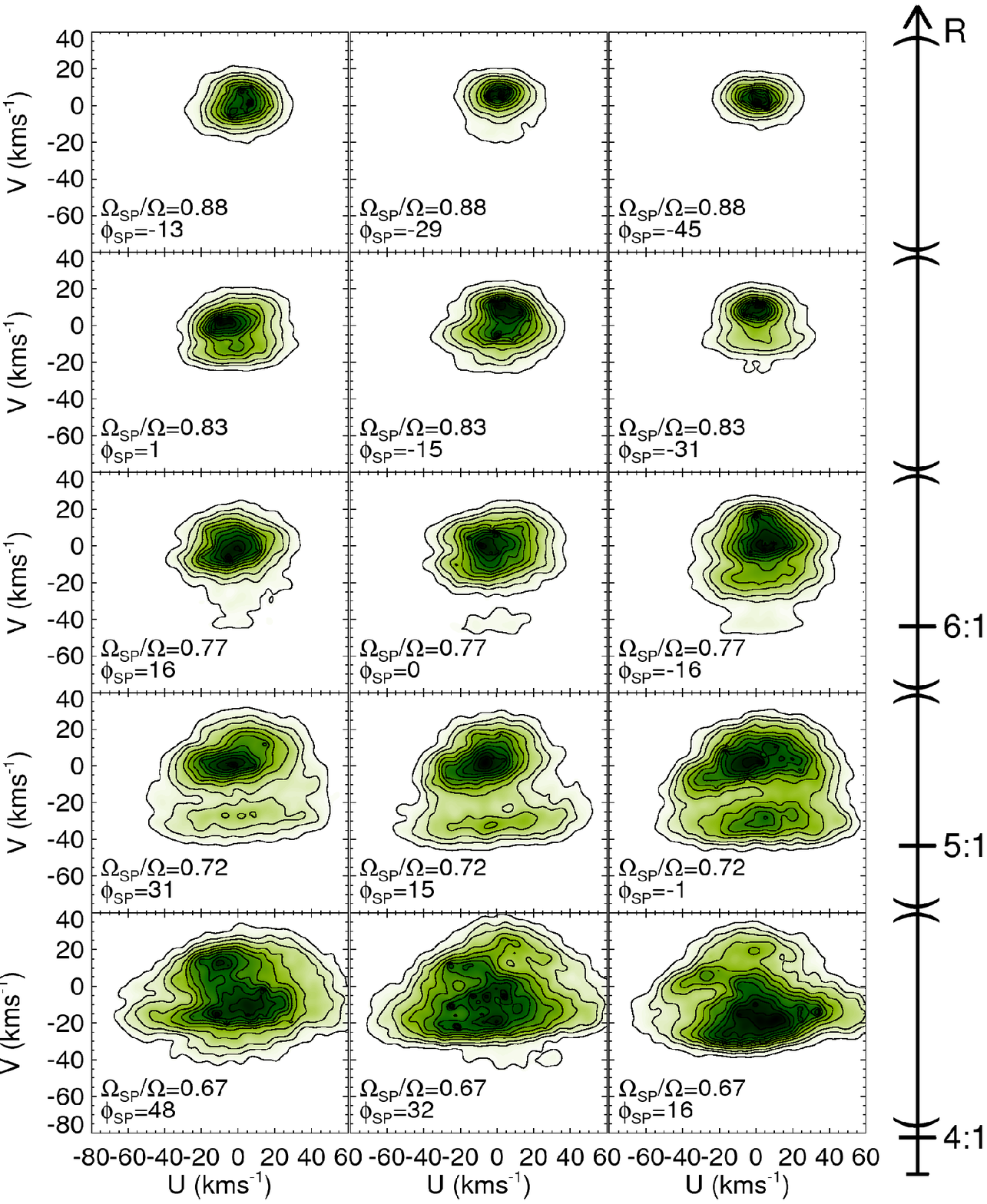}
}
\caption{$U-V$ velocity distributions for the 15 regions of Fig. \ref{fig:1} around the Sun (left) and close to a spiral arm (right) for the
simulations with the \name model (Table \ref{tab:1}). The 15 panels are positioned
in the figure similarly to their spatial location of Fig. \ref{fig:1}. The scale in the
right indicates the spatial position of the spiral resonances.}
\label{fig:2}       
\end{figure}

Here we present one example of the simulations corresponding to the parameters listed in Table \ref{tab:1}. The locus of the spiral structure is shown in Figure \ref{fig:1}. For this simulation we show the velocity distributions of two sets of 15 regions. One of them is located around the Sun and the other is near a spiral arm position (Figure \ref{fig:1}). They have $300\pc$ of radius, and are separated by $600 \pc$ in galactocentric radius and $\sim2 \kpc$ in azimuth. The results are shown in Figure \ref{fig:2} for the regions around the Sun (left) an near the spiral arms (right). This figure and the rest of the simulations show that the models predict a strong stellar kinematic response to the spiral arms highly dependent on disc position. The kinematic groups induced by the spiral arms change significantly if one moves only $\sim0.6\kpc$ in galactocentric radius. Changes in azimuth have a larger scale of $\sim2\kpc$. For all models and for all density contrasts, within the observational MW range, the spiral arms induce strong imprints close to the 4:1 inner resonance. By contrast, there is almost no structure at corotation. Depending on the spiral pattern speed, which determines the position of the 4:1 resonance, the kinematic imprints are more intense in nearby vicinities or far from the Sun. The particular example shown in  Figure \ref{fig:2} also indicates that there are more and stronger groups as one approaches the spiral arms. 

\section{RAVE: Observed kinematic groups across the disc}
\label{sec:2}

The RAVE (RAdial Velocity Experiment, \cite{Steinmetz06}) spectroscopic survey provides the means to make a giant leap forward in the understanding of the origin of the moving groups. It has already collected more than 500,000 spectra giving radial velocity precisions of $\sim2\kms$. \cite{Breddels10}, \cite{Zwitter10} and \cite{Burnett11} have derived spectrophotometric distances for part of the sample. Part of the RAVE stars have also now been cross-matched with proper motions data, giving, together with the distances, the 6D phase space distribution for a sample of 203025 stars. The data for the solar neighbourhood from this sample is already larger in number than previously available kinematic data for the local volume, such as the Geneva-Copenhagen survey (GCS). For instance, there are 31524 stars enclosed in a sphere around the Sun of $200 \pc$ compared to the 13491 in the GCS. But more importantly, the volume sampled by the RAVE stars is much larger, giving kinematics representative of distant regions of the thin and thick disc. For instance, out of these 203025 stars, 112792 stars have height above/below the plane $|Z|\leq300\pc$ and extend at least up to $2\kpc$.

Here we present a preliminary analysis of the RAVE 6D phase space sample by analysing the local velocity distribution (sphere around the Sun of $200\pc$). To detect the kinematic substructures in velocity distribution we use a technique based on the Wavelet transform (WT) and the {\em \`a trous} (``with holes'') algorithm \cite{Starck02}. The WT yields, for every position in the velocity plane, coefficients equal to 0 for a constant signal and positive values for overdensities for a certain desired substructure scale. The significance of the structures, biases, etc, will be discussed in Antoja et al. in prep.

\begin{figure}
\centering
\resizebox{0.9\columnwidth}{!}{%
 \includegraphics{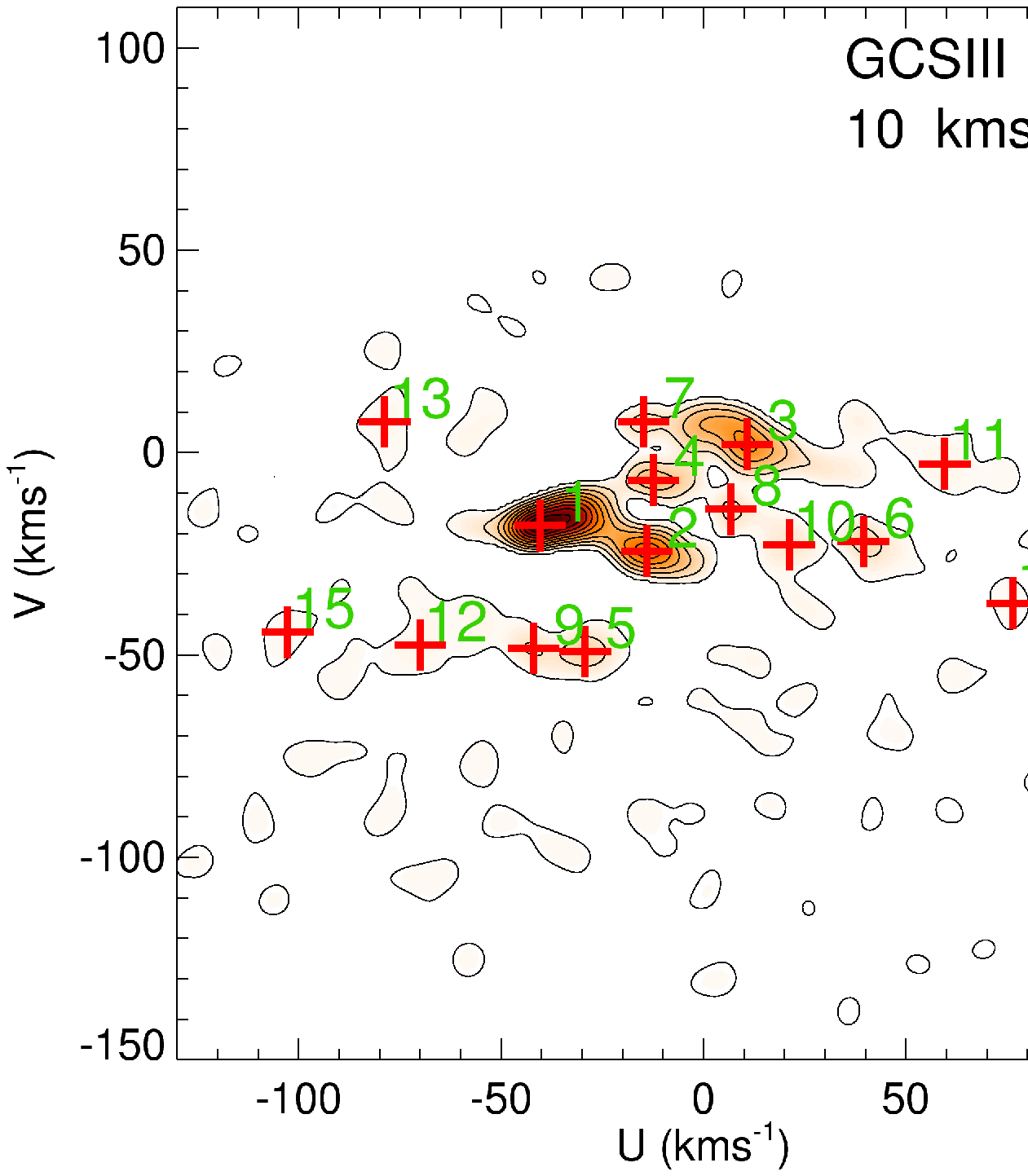} 
 \includegraphics{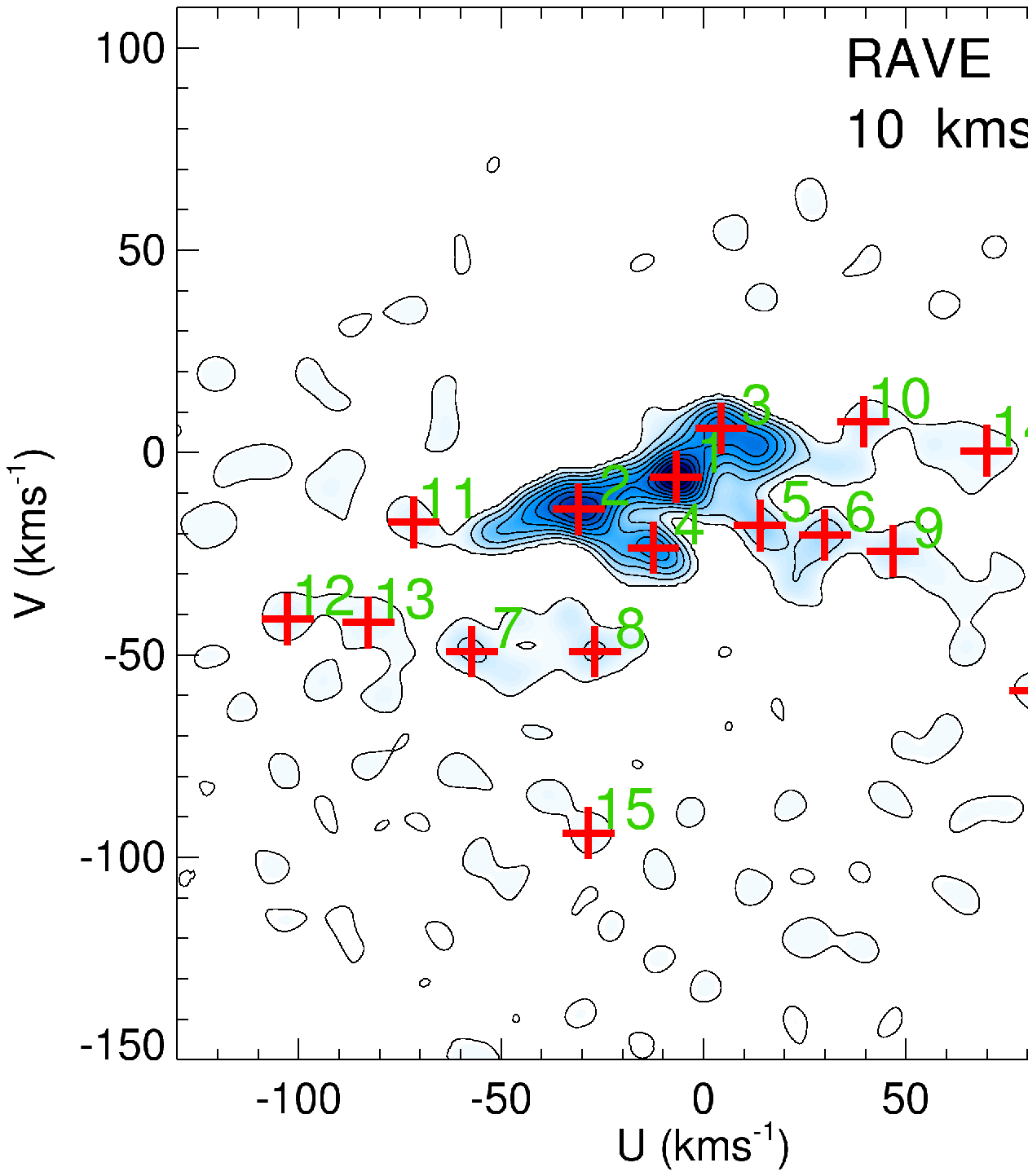}
}
\caption{Velocity structures at scales of $10\kms$ in the solar neighbourhood for the GCSIII (left) and for RAVE (right). Red crosses show some of the maxima of the detected structures.}
\label{fig:3}       
\end{figure}

Figure \ref{fig:3} (left) shows the velocity structures for the  new version of the GCS (GCSIII) by \cite{Holmberg09} (left), and for the RAVE local sample (right).The local velocity distribution revealed by the WT is dominated by clear overdensities. Many structures are elongated in $V$ and slightly tilted. The groups that we find in the two independent samples are basically the same and similar to the ones found in previous studies with other samples (e.g. \cite{Dehnen98,Chereul99,Antoja08}). For instance, the four main groups that we find in both samples (1 to 4, but in different order) are the well-known kinematic groups of Coma Berenice, Hyades, Sirius and Pleiades. Also Hercules is well-detected as an elongated structure in the $U$ direction and at $V\sim-50\kms$. RAVE samples of larger spatial extent ($500 \pc$ radius) centred in the Sun show basically the same kinematic features.

\section{Conclusions}\label{conc}

The models predict a stellar kinematic response to the spiral arms strongly dependent on disc position. The strongest effects are seen near the 4:1 resonance. By detecting observationally the direction of maximum kinematic substructure, we may determine the radius of this resonance and, therefore, the spiral arm pattern speed. The RAVE sample allows us, for the first time, to study the dependence on Galactic position of the (thin and thick) disc moving groups (Antoja et al., in prep.). This can be used to test some of the trends noticed in our models. A decisive point will be reached when the comparison between models and observed velocity distribution at different disc positions with RAVE and the forthcoming Gaia data will allow us to constrain the non-axisymmetries of the Galaxy disc. A solar neighbourhood sample from the RAVE catalogue shows the expected kinematic groups but now with more statistics and over a larger spatial volume. RAVE samples of larger spatial extent centred in the Sun show basically the same kinematic groups, indicating they are large scale features, and pointing out, once more, that they are indeed dynamical effects.





\end{document}